\documentclass[12pt]{article}
\usepackage{times}
\usepackage{graphicx}
\usepackage{color}
\usepackage{multirow}
\usepackage{rotating}
\usepackage{bbm}
\usepackage{latexsym}
\usepackage{doi}
\usepackage{hyperref}

\providecommand\doi[1]{\href{https://doi.org/#1}{\url{#1}}}
\usepackage{amsmath}
\usepackage{amsthm}
\usepackage{amsfonts}
\usepackage{amssymb} 
\usepackage{graphicx}
\usepackage{rotating}

\usepackage[ruled,vlined]{algorithm2e}

\newtheorem*{theorem*}{Theorem}

\usepackage{xcolor}

\newcommand{\captionfonts}{\normalsize}

\makeatletter  
\long\def\@makecaption#1#2{%
  \vskip\abovecaptionskip
  \sbox\@tempboxa{{\captionfonts #1: #2}}%
  \ifdim \wd\@tempboxa >\hsize
    {\captionfonts #1: #2\par}
  \else
    \hbox to\hsize{\hfil\box\@tempboxa\hfil}%
  \fi
  \vskip\belowcaptionskip}
\makeatother   

\begin{document}

\ \vspace{-10mm}\\{\Large Cheap robust learning of data anomalies with analytically-solvable entropic outlier sparsification }
\begin{center}

{\bf \large  Illia Horenko$^{\displaystyle 1}$}\\
{$^{\displaystyle 1}$Universit\'{a} della Svizzera Italiana (USI), Institute of Computing, Faculty of Informatics, Via G. Buffi 13, TI-6900 Lugano, Switzerland}\\
\end{center}


.     

\markboth{}{NC instructions}
%
%
{\bf Abstract: }
Entropic Outlier Sparsification (EOS) is proposed as a cheap and robust computational strategy for the detection of data anomalies in a broad class of learning methods, including the unsupervised problems (like detection of non-Gaussian outliers in mostly-Gaussian data) and in the supervised learning with mislabeled data.  EOS dwells on the derived analytic closed-form solution of the (weighted) expected error minimization problem subject to the Shannon entropy regularization.  In contrast to common regularization strategies requiring computational costs that scale polynomial with the data dimension, identified closed-form solution is proven to impose additional iteration costs that depend linearly on statistics size and are independent of data dimension. Obtained analytic results also explain why the mixtures of spherically-symmetric Gaussians - used heuristically in many popular data analysis algorithms - represent an optimal choice for the non-parametric probability distributions when working with squared Euclidean distances, combining expected error minimality, maximal entropy/unbiasedness, and a linear cost scaling. The performance of EOS is compared to a range of commonly-used tools on synthetic problems and on partially-mislabeled supervised classification problems from biomedicine. 
%
%
\newpage

Detection of data anomalies, outliers, and mislabeling is a long-standing problem in statistics, machine learning (ML), and artificial intelligence \cite{donoho1982breakdown,rocke1996identification,filzmoser2008outlier,wang2019}. Let $\left\{x_1,x_2,\dots,x_T\right\}$ be a fixed dataset (where data instances $x_t$ are possibly augmented with labels), $\theta$ be a set of ML model parameters and $g(x_t,\theta)$ be a scalar-valued error function measuring a misfit of the data instance $x_t$. Then, a wide class of learning methods and anomaly detection algorithms can be formulated as numerical procedures for a minimization of the following functional:   
\begin{eqnarray}\label{eq:functional_orig}
	 \left\{\hat{w},\hat{\theta}\right\}&=&\arg\min\limits_{w_{\cdot},\theta}\sum_{t = 1}^T w_t g(x_t,\theta),
\end{eqnarray}
where $0<w_t<1$ is the \emph{outlyingness}, taking the values close to zero if the  data point $x_t$ is an anomaly \cite{stahel1981robuste,donoho1982breakdown,rousseeuw1990unmasking,maronna1995behavior}.
If $w$ and $\theta$ are both unknown then the above problem (\ref{eq:functional_orig}) for simultaneous estimation of model parameters and error weights  becomes ill-posed. Common approaches deal with this ill-posedness problem imposing additional parametric assumptions on $w$, e.g., based on parametric thresholding of one-dimensional linear projections in Stahel-Donoho estimators or deploying other parametric tools (like $\chi(D)$-distribution quantiles to determine outliers of a D-dimensional normal distribution)  \cite{stahel1981robuste,donoho1982breakdown,maronna1995behavior,zuo2004stahel}. An appealing idea would be to make this ill-posed problem well-posed in a nonparametric way, by regularizing it with one of the common regularization approaches. For example, applying $l1$-regularization could result in a sparsification of $w$ and zeroing-out the outlying data points from the estimation \cite{donoho2006most}. However, applying $l1$ and other sparsification methods results in a polynomial cost scaling required for a numerical solution of the resulting optimization problems - and would limit the solution of  (\ref{eq:functional_orig}) to relatively small problems \cite{huang19}.

The key message of this brief report is in showing that the simultaneous well-posed detection of anomalies and learning of parameters $\theta$ in (\ref{eq:functional_orig})  can be achieved computationally very efficiently by means of the minimization of expected errors from the right-hand side of (\ref{eq:functional_orig})  - performed simultaneously to the regularized Shannon entropy maximization of the error weight distribution $w$:  
\begin{eqnarray}\label{eq:functionalT1}
	 \left\{w^{(\alpha)},\theta^{(\alpha)}\right\}&=&\arg\min\limits_{w \in \mathbb{P}^{(T)}} L(w,\theta,\alpha), \nonumber\\
	  \textrm{where} ~L(w,\theta,\alpha) &=& \sum_{t = 1}^T w_t g(x_t,\theta) + \alpha\sum_{t = 1}^T w_t \log w_t,\nonumber\\
	 \textrm{such that } ~~ w&\in&\mathbb{P}^{(T)},\nonumber\\ 
	\mathbb{P}^{(T)} &:=& \left\{ w\in \mathbb{R}^{T} \bigg | \; w \geq 0\; \wedge \; \sum_{t=1}^T w_{t}=1 \right\},
\end{eqnarray}
The following Theorem summarizes the properties of this problem's solutions: 
\begin{theorem*}\label{theorem:analytic_solution}
For any fixed $\left\{x_1,x_2,\dots,x_T\right\}$  and $\theta$ such that  $\sup_t |g(x_t,\theta)|<\infty$ and $\alpha>0$, constrained minimization problem (\ref{eq:functionalT1})
admits a unique closed-form solution $w^{(\alpha)}$:
\begin{eqnarray}\label{eq:sol}
	w^{(\alpha)}_t &=& \frac{\exp\left( - \alpha^{-1} g(x_t,\theta)\right)}{\sum_{t=1}^T\exp\left( - \alpha^{-1} g(x_t,\theta) \right)}\;. 
\end{eqnarray}
\end{theorem*}
The proof of the Theorem is provided in the Supplemental Information available at \url{https://www.dropbox.com/s/6psesbsgcckfh1p/EOS_supplement.pdf?dl=0}. It is straightforward to validate that the numerical cost of computing (\ref{eq:sol}) scales linearly in statistics size $T$ and is independent of the data dimension $D$ - in contrast to common regularization techniques that require cost scaling polynomial in the data dimension $D$  \cite{huang19}.

 If the error function $g(x_t,\theta)$ is a squared Euclidean distance (as in the least-squares methods) then according to the above Theorem the unique probability distributions $w$ minimizing the  (\ref{eq:functionalT1}) are from the $\alpha$-parametric family of spherically-symmetric Gaussians, with the dimension-wise variance $\sigma^2$ being $\sigma^2=0.5\alpha$.   This result provides an interesting insight into the density-based methods, for example in the t-Stochastic Neighbour Embedding, t-SNE \cite{maaten08} - one of the most popular nonlinear dimensions reduction approaches in the area of biomedicine (with over 20'000 citations according to Google Scholar). This method searches for the optimal low-dimensional approximations of the high-dimensional densities  defined in a heuristic way as mixtures of spherically-symmetric Gaussians
 \begin{eqnarray}\label{eq:SSG}
w_t= \frac{\exp\left( - \||x_i-x_j \||^2/2\sigma^2\right)}{\sum_{k}\exp\left( - \||x_i-x_k \||^2/2\sigma^2\right)}
\end{eqnarray}
 with a multiindex $t=(i,j)$. According to the above Theorem, this heuristics - building a computational foundation of tSNE - is actually equivalent to the most optimal nonparametric density estimate  (\ref{eq:sol}), in a sense that it is simultaneously minimizing the expectation of the pairwise squared Euclidean distances between the data points  (when considering  $g(x_t,\theta)=\||x_i-x_j \||^2$ as an error-function in (\ref{eq:functionalT1})), simultaneously maximizing the entropy of $w$ (i.e., providing the least-biased estimation) and is obtained with an explicitly-computable closed form expression. Furthermore, solution (\ref{eq:sol}) also provides a recipe for computing such tSNE density estimates in the cases with non-Euclidean error functions $g$.   
 
  \begin{algorithm}
\small
\SetAlgoLined
\KwResult{Optimal values of $w$ and $\theta$ minimizing the functional $L$ in \eqref{eq:functionalT1}}
 For given  $\left\{x_1,x_2,\dots,x_T\right\}$, and  $\alpha>0$, randomly choose initial $w^{(1)}$\;
 $I=1;\, L^{(I)}=\infty;\, \Delta L^{(I)}=\infty$\;
 \While{$\Delta \mathcal{L}^{(I)}>tol$}
 {\underline{$\theta$-step}: find $\theta^{(I)}$ as a solution of  \eqref{eq:functionalT1} for fixed  $w^{(I)}$\;
	\underline{$w$-step}: find $w^{(I+1)}$ evaluating the explicit expression (\ref{eq:sol})  for fixed  $\theta^{(I)}$\;
	$L^{(I+1)}= L(w^{(I+1)},\theta^{(I)},\alpha)$\;
	$I = I + 1$\;
	$\Delta L^{(I)}= L^{(I-1)}-L^{(I)}$\; 
 }
 \caption{EOS algorithm for the Solution of Optimization Problem (\ref{eq:functionalT1})}\label{alg:eSPAgeneric}
\end{algorithm}
\normalsize
 \begin{figure}
       \includegraphics[width=1\textwidth]{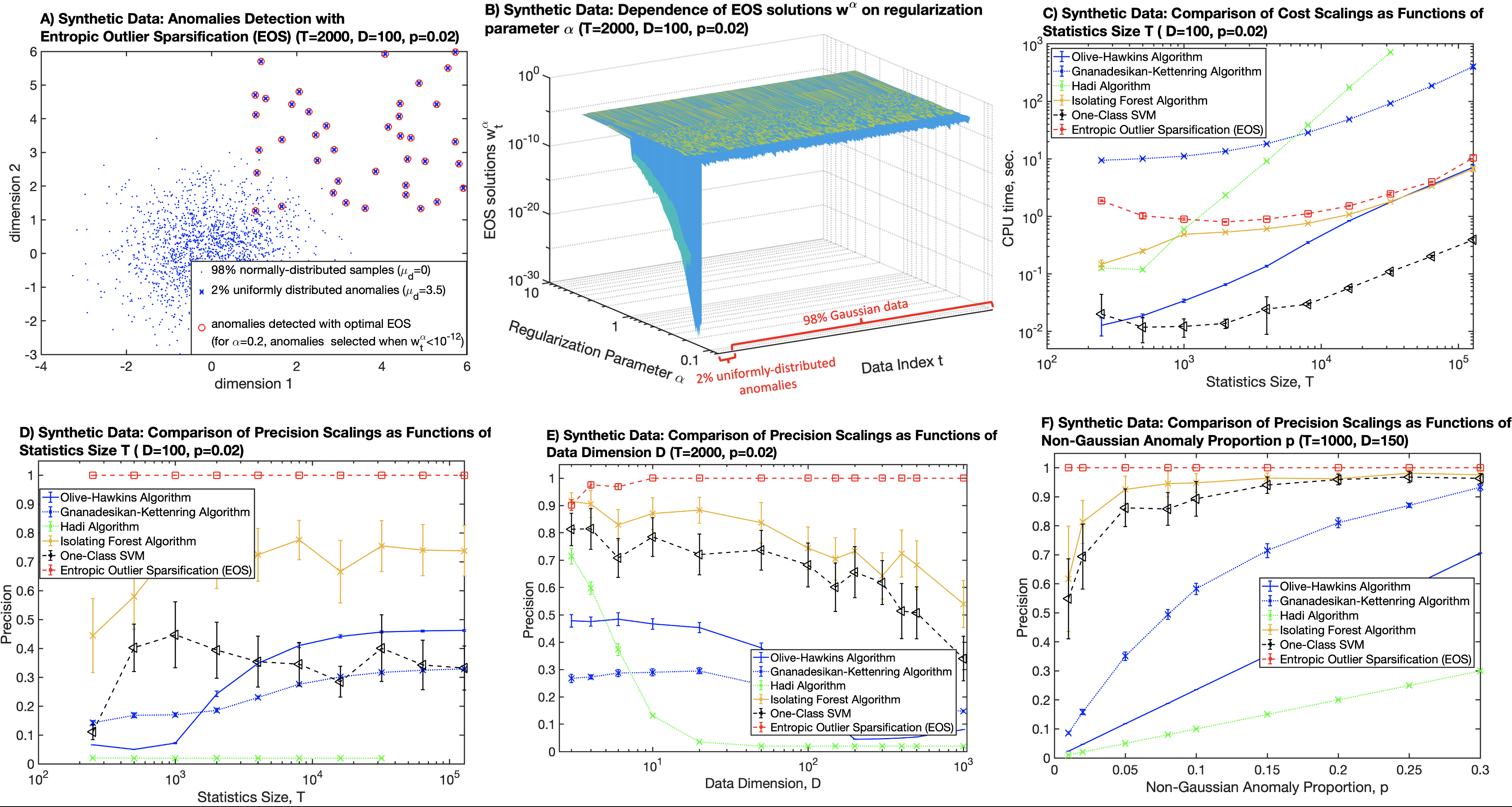}
       \includegraphics[width=1\textwidth]{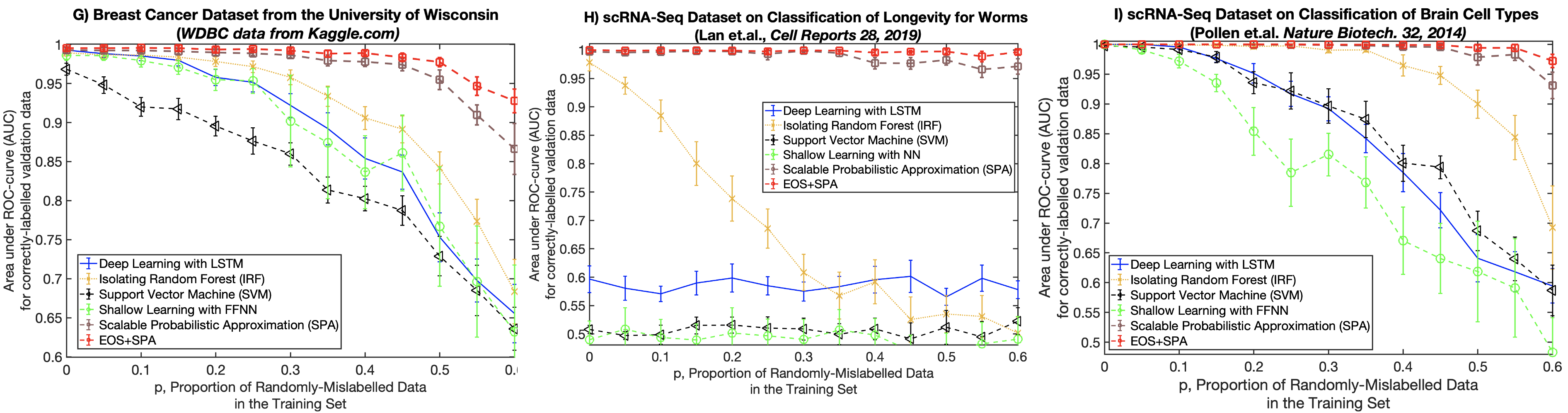}
 \caption{Comparison of EOS algorithm for the solution of optimization problem (\ref{eq:functionalT1}) to common methods of data anomaly detection (A-F) and supervised classifier learning (G-I) on synthetic and real data examples.  \textbf{A-F}: Synthetic data examples: unsupervised detection of uniformly-distributed anomalies in the Gaussian data. \textbf{G-I}: Real data examples from biomedicine: multiple cross-validated comparisons for mislabeled training data. }\label{fig:results}
\end{figure}

 It is straightforward to verify that the simultaneous learning of the parameters  $\theta$ and probability densities $w$ can be performed with the monotonically-convergent Entropic Outlier Sparsification (EOS) algorithm (see Algorithm 1). 

Fig. 1 summarizes numerical experiments comparing EOS to common data anomaly detection and learning tools on randomly-generated synthetic datasets (representing multivariate normal distributions with asymmetrically positioned uniformly-distributed outliers, see Figure 1 panels A-F) and three biomedical data sets with various proportions of randomly-mislabeled data instances in the training sets (see Figure 1 panels G-I). All of the compared algorithms are provided with the same information and run with the same hardware and software, 50 cross-validations were performed in every experiment to visualize the obtained 95\% confidence intervals. In numerical experiments with synthetic data (panels A-F) EOS algorithm is deployed with $g$ being the negative point-wise multivariate Gaussian loglikelihood, i.e. with $g(x_t,\mu,\Sigma)=1/D\left(0.5*\log(\det(\Sigma))+0.5(x_t-\mu)^*\Sigma^{-1}(x_t-\mu)\right)$, where $\mu$ and $\Sigma$ are Gaussian mean and covariance, respectively.  Iterative estimation of weighted mean and covariance in the $\theta$-step of EOS algorithm is performed using analytical estimates of the weighted Gaussian covariance and mean, convergence tolerance $tol$ is set to $10^{-12}$. Total computational costs and statistical precisions - the latter are measured as the numbers of correctly identified points not belonging to the Gaussian distribution divided by the total number of identified outliers - are performed for various problem dimensions, statistic sizes, and outlier proportions. EOS was compared to all of the outlier detection methods available in the "Statistics" and "Machine Learning" toolboxes of MathWorks. Precision is chosen as measure of performance here since it is more robust then the other common measures when the  data sets are not balanced - e.g., when the number of instances in one class (outliers) is much less then in the other class (non-outliers). These results show that EOS allows a marked and robust improvement of outlier detection precision for all of the considered comparison cases.

Next, real labeled data sets from the biomedicine are considered, including the popular WDBC breast cancer diagnostics data set from \url{Kaggle.com} (panel G) and two single-cell mRNA (scRNA-Seq) gene expression data sets from longevity research \cite{Lan19} (panel I) and from neuroscience \cite{pollen14} (panel H). The main focus here is on comparing the robustness of learning methods to randomly mislabeled data instances in the training set, for common binary classifiers and for EOS that is equipped with error function $g$ from the Scalable Probabilistic Approximation (SPA) classifier algorithm \cite{Gerber_2020,horenko_2020}.  SPA is selected since it shows the highest robustness to mislabeling for all of the considered data sets (see panels G-I). As can be seen from the Figure, EOS with $g(x_t,\theta)$ from SPA (EOS+SPA, red dashed lines), allows a statistically-significant improvement of prediction performance (measured with the common performance measure AUC) for all of the considered mislabeling proportions $p$ for all three of the considered biomedical examples.

 The code reproducing these results is provided for open access at \url{https://www.dropbox.com/sh/zqio1zdrlat368j/AAD0s5mjvheEO8O2gjSAJ4ila?dl=0}.
 \newpage
\bibliographystyle{unsrt}
\bibliography{biblio}

\end{document}